\documentclass[12pt]{article}
\usepackage{geometry}
\usepackage{a4}
\usepackage{graphicx}
\usepackage{epsf}
\usepackage{amsmath}
\usepackage{amssymb}
\usepackage{cite}
\newcommand{\be}{\begin{equation}}
\newcommand{\ee}{\end{equation}}

\newcommand{\Rmnum}[1]{\expandafter\@slowromancap\romannumeral #1@}
\newcommand{\bea}{\begin{eqnarray}}
\newcommand{\eea}{\end{eqnarray}}

\begin{document}
\def\A{{\mathbb{A}}}
\def\B{{\mathbb{B}}}
\def\C{{\mathbb{C}}}
\def\R{{\mathbb{R}}}
\def\s{{\mathbb{S}}}
\def\T{{\mathbb{T}}}
\def\Z{{\mathbb{Z}}}
\def\W{{\mathbb{W}}}
\begin{titlepage}
\title{Geodesics in Information Geometry : Classical and Quantum Phase Transitions}
\author{}
\date{
Prashant Kumar, Subhash Mahapatra, Prabwal Phukon, \\ Tapobrata Sarkar
\thanks{\noindent E-mail:~ kprash, subhmaha, prabwal, tapo @iitk.ac.in}
\vskip0.4cm
{\sl Department of Physics, \\
Indian Institute of Technology,\\
Kanpur 208016, \\
India}}
\maketitle
\abstract{
\noindent
We study geodesics on the parameter manifold, for systems exhibiting second order classical and quantum phase transitions. The coupled non-linear
geodesic equations are solved numerically for a variety of models which show such phase transitions, in the thermodynamic limit. 
It is established that both in the classical as well as in the quantum case, geodesics are confined to a single phase, and exhibit turning behavior near 
critical points. Our results are indicative of a geometric universality in widely different physical systems.
}
\end{titlepage}

\section{Introduction}

Information theoretic studies of phase transitions are, by now, well established. The underlying idea here is geometric in nature, and rests
on the definition of a Riemannian metric tensor on the space of parameters (called parameter manifold) of a system. Depending on whether the interactions of 
the system are classical or quantum in nature, this metric might be induced from the equilibrium thermodynamic state space \cite{rupp1}
(for a review, see \cite{rupp}), or from the natural Hilbert space structure of quantum states \cite{pv}. For the former, the parameter manifold consists of thermodynamic 
control parameters such as the pressure, volume and temperature, while for the latter, this might be thought of as the manifold of coupling constants
appearing in the Hamiltonian.

Given such a metric tensor, the parameter manifold can have very different properties depending on whether the system undergoes 
a second order classical or a quantum phase transition (CPT or QPT). Whereas the hallmark of a CPT is that the scalar curvature arising out of the metric 
diverges at a second order phase transition (and everywhere on the spinodal curve), this is not the case for second order QPTs where the
curvature can remain regular \cite{zan1}. 
It is also known that whereas some components of the metric tensor vanish at a second order CPT, as these are related to inverses of
thermodynamic response coefficients \cite{rupp}, for QPTs, the situation is reversed, and some of
the components of the metric tensor  diverge at such a transition, as follows from first order perturbation theory \cite{zan1} (although this may not be true in some 
special cases, see \cite{zan1a}).

Although a lot of attention has been paid to the behavior of the metric tensor and its associated scalar curvature in the context of phase transitions,
much less is known about geodesics, i.e paths that minimize the distance between two points on the parameter manifold. In any geometric setup, 
the behavior of geodesics is an important object to study. Some studies on geodesics have appeared in the context of CPTs 
\cite{diosi}, and QPTs (specifically, for adiabatic quantum computation) \cite{zan2},\cite{zan3}, in special cases.  The purpose of this paper is to complement and
generalize these results, and to obtain and analyze general solutions to the geodesic equations for some model systems that exhibit 
second order phase transitions. 

Here, we study four models in the thermodynamic limit :
the Van der Waals (VdW) model for fluids, the Curie Weiss (CW) mean field model of ferromagnetism,  the infinite Ising ferromagnetic chain - all of which exhibit 
CPTs at finite temperature, and the transverse field XY model that exhibits a QPT at zero temperature. For all these models, the full set of coupled non-linear 
geodesic equations in the information geometric context are set up and solved numerically, with appropriate initial conditions. 
To the best of our knowledge, such an analysis has not been performed before. 
Our treatment is completely general in nature, and differs significantly from the methods used in
\cite{diosi},\cite{zan3} where the focus was on obtaining specific geodesics between two given points in the parameter manifold. Interestingly, we find that 
in all the examples that we consider, geodesics exhibit a ``turning point'' close to criticality, 
and are ``confined'' to a single phase, thus indicating a geometric universality in apparently unrelated physical phenomena. 

This paper is organized as follows. In the next section, we first briefly recall some basic facts about information geometry and geodesics. We then proceed
to analyze the VdW, the CW and the infinite Ising ferromagnet, as 
illustrations of CPTs. For the Ising ferromagnet, our analysis of information geometry is novel, and has not appeared in the literature before. 
In section 3, we analyze the geodesic structure of QPTs via the transverse field XY spin chain. We end in section 4 with our discussions
and directions for future study. 

\section{Information Geometry, Geodesics, and Classical Phase Transitions}

In the context of equilibrium thermodynamics of classical systems, the formulation of information geometry is mainly due to the work of Ruppeiner \cite{rupp}. 
The main idea here is to consider the positive definite Riemannian metric arising out of the Hessian of the entropy density $s$, and given by a line element
\begin{equation}
d\tau^2 = g_{\mu\nu}dx^{\mu}dx^{\nu} ~~~g_{\mu\nu} = -\frac{1}{k_B}\left(\frac{\partial^2 s}{\partial x^{\mu} \partial x^{\nu}}\right)
\label{entropyrep}
\end{equation}
Here, $x^{\mu}, \mu=1,2$, denotes the internal energy and the particle number per unit volume, and are co-ordinates on the parameter manifold 
in the ``entropy representation.''  $k_B$ is the Boltzmann's constant, which we will set to unity in what follows. 
The line element of eq.(\ref{entropyrep}) introduces the concept of a distance in the space of equilibrium thermodynamic states via fluctuation theory, 
i.e, the larger is this distance between two given states, the smaller is the probability that these are related by a thermal fluctuation. Various representations
(related to each other by Legendre transforms) can be used for this geometric construction (a full list can be found in \cite{rupp}), and a particularly useful 
diagonal form of the metric for single component fluids and magnetic systems is 
\begin{equation}
d\tau^2 = \frac{1}{T}\left(\frac{\partial s}{\partial T}\right)_{\rho} dT^2+ \frac{1}{T}\left(\frac{\partial \mu}{\partial \rho}\right)_Td\rho^2
\label{line}
\end{equation}
where $T$ is the temperature, $\rho$ the number density, and $\mu = \left(\frac{\partial f}{\partial \rho}\right)_T$, $f$ being the Helmholtz free energy per unit volume. 
For magnetic systems, we need to consider thermodynamic quantities per unit spin, with the magnetization per unit spin $m$ replacing $\rho$. 

On the other hand, information geometry in quantum mechanical systems, first studied by Provost and Vallee \cite{pv}, is defined by taking two 
infinitesimally separated quantum states and constructing the quantity
\begin{equation}
|\psi\left({\vec x} +d{\vec x}\right) - \psi\left({\vec x}\right)|^2 = \langle \partial_{\mu} \psi|\partial_{\nu} \psi\rangle dx^{\mu}dx^{\nu} = \alpha_{\mu\nu}dx^{\mu}dx^{\nu}
\label{pv1}
\end{equation}
where $x^{\mu}$ (collectively denoted as ${\vec x}$ in the l.h.s of eq.(\ref{pv1})) denotes the parameters on which the wave function $\psi$ depends on, and
$\partial_{\mu}$ is a derivative with respect to $x^{\mu}$. 
From the $\alpha_{\mu\nu}$ (which are not gauge invariant), a meaningful gauge-invariant metric tensor can be defined as \cite{pv}
\begin{equation}
g_{\mu\nu} = \alpha_{\mu\nu} - \beta_{\mu}\beta_{\nu};~~~~\beta_{\mu} = -i\langle\psi\left({\vec x}\right)|\partial_{\mu}\psi\left({\vec x}\right)\rangle
\label{metric}
\end{equation}
Here, $g_{\mu\nu}$ is the metric induced from the natural structure of the Hilbert space of quantum states. 
The metrics in eqs.(\ref{line}) and (\ref{metric}) can be used to predict second order phase transitions in both CPTs \cite{rupp} and QPTs \cite{zan1}. We also
record here the expression for the scalar curvature arising out of the metric in the special case when the metric is diagonal (with $g \equiv {\rm det}~g_{\mu\nu}$) :
\begin{equation}
R = \frac{1}{\sqrt{g}}\left[\frac{\partial}{\partial x^1}\left(\frac{1}{\sqrt{g}} \frac{\partial g_{22}}{\partial x^1}\right) + 
\frac{\partial}{\partial x^2}\left(\frac{1}{\sqrt{g}} \frac{\partial g_{11}}{\partial x^2}\right)\right]
\label{scalarcurvature}
\end{equation}

Given the information geometry of classical or quantum systems, we wish to study geodesics in the same. Let us briefly recall a few elementary facts about geodesics. 
For a manifold endowed with a metric with components $g_{\mu\nu}$, a geodesic is a path that extremizes the proper 
distance (or line element, whose infinitesimal form is given by $d\tau^2 = g_{\mu\nu}dx^{\mu}dx^{\nu}$). This can be cast as a variational problem, to determine the extrema of 
the integral $\int_1^2 \sqrt{g_{\mu\nu}{\dot x^{\mu}}{\dot x^{\nu}}}d\lambda$
where the dot denotes a derivative with respect to $\lambda$, which is an affine parameter, parametrizing the curve joining two points denoted $1$ and $2$. 
Calculus of variations can then be applied with the result that geodesic curves are solutions to the differential equations 
\begin{equation}
{\ddot x^{\mu}} + \Gamma^{\mu}_{\nu\rho}{\dot x^{\nu}}{\dot x^{\rho}} = 0,~~~
{\rm with}~~~\Gamma^{\mu}_{\nu\rho} = \frac{1}{2}g^{\mu\zeta}\left(\frac{\partial g_{\zeta\nu}}{\partial x^{\rho}} + \frac{\partial g_{\zeta\rho}}{\partial x^{\nu}}
- \frac{\partial g_{\nu\rho}}{\partial x^{\zeta}}\right)
\label{geodesic}
\end{equation}
The above equation can also be obtained by writing a ``Lagrangian'' 
\begin{equation}
{\mathcal L} = \frac{1}{2}\left(g_{\mu\nu}{\dot x^{\mu}}{\dot x^{\nu}}\right)
\label{lagrangian}
\end{equation}
and using the (derivatives of the) Euler-Lagrange equations that follow. This method often provides valuable insights into the symmetries of the system. 
We will be interested in studying the solutions of eq.(\ref{geodesic}) in the context of CPTs and QPTs. 
It is well known that a natural affine parameter for geodesic curves is $\lambda = \tau$, and thus it is useful to consider the normalized vector 
${\dot x^{\mu}} = dx^{\mu}/d\tau$ such that ${\dot x^{\mu}}{\dot x_{\mu}} = g_{\mu\nu}{\dot x^{\mu}}{\dot x^{\nu}} = 1$. 

Eq.(\ref{geodesic}), in general, gives rise to a set of coupled non-linear differential equations, which might be difficult to solve analytically. 
We will mostly concentrate on numerical solutions with appropriate boundary conditions. Note that in terms of the normalized
vector ${\dot x^{\mu}}$, we need to specify three boundary conditions in order to solve eq.(\ref{geodesic}), with the fourth one being fixed by the 
normalization condition. Namely, we choose a ``starting point,'' i.e, an initial value of $x^{\mu}$, and any one component of ${\dot x}^{\mu}$. The second 
component of the derivative is then determined from the fact that ${\dot x^{\mu}}$ is normalized. 
With these boundary conditions, we determine the most general solutions to eq.(\ref{geodesic}), and study geodesics near criticality.
This is done by solving for $x^{\mu}$ in terms of the affine parameter $\tau$, and tracing out the geodesic near the critical point, 
by parametrically plotting the resulting solution, under variation of $\tau$. 
\footnote{We will also keep in mind that geodesic paths are not unique : an elementary example is that of a 2-sphere, where there are an infinite number of
geodesics, i.e great circles, between two anti-podal points.}
Let us now illustrate the above discussion with the example of the Van der Waals fluid and the Curie-Weiss ferromagnet. 

\subsection{The Van der Waals and the Curie-Weiss Models}

Information geometry of the Van der Waals fluid is well established, see e.g \cite{brodyhook}. We start from the Helmholtz free energy per unit volume
\begin{equation}
f_{{\rm VdW}} = -\rho T{\rm ln}\left(\frac{e}{\rho}\right) + \rho c_v T{\rm ln}\left(\frac{e}{T}\right) - \rho T{\rm ln}\left(1 - b\rho\right) - a\rho^2
\end{equation}
where $e$ denotes the exponential function, $c_v$ is the specific heat at constant volume, $\rho$ and $T$ are the number density per molecule of fluid and 
the temperature, and $a$, $b$ are the coefficients arising in the VdW equation of state. 
It is convenient to work with the reduced VdW equation of state, and we can substitute $a = 9T_c/8\rho_c$, $b = 1/3\rho_c$, where $\rho_c$ and $T_c$
denote the critical values of the density and the temperature, respectively. Further, the reduced
density and temperature are defined by $\rho_r = \rho/\rho_c$, $T_r = T/T_c$. We will set $\rho_c = T_c = 1$ to simplify the algebraic details, and also
choose $c_v = 3/2$, the ideal gas value. The information metric (in terms of the co-ordinates $T_r$ and $\rho_r$) is then given, from eq.(\ref{line}), by 
\begin{equation}
g_{TT} = \frac{3}{2} \frac{\rho_r}{T_r^2},~~~~
g_{\rho\rho} = \frac{9\left[4T_r - \rho_r\left(\rho_r - 3\right)^2\right]}{4\rho_rT_r\left(\rho_r - 3\right)^2}
\label{vdwfull}
\end{equation}
\begin{figure}[t!]
\centering
\includegraphics[width=2.8in,height=2.3in]{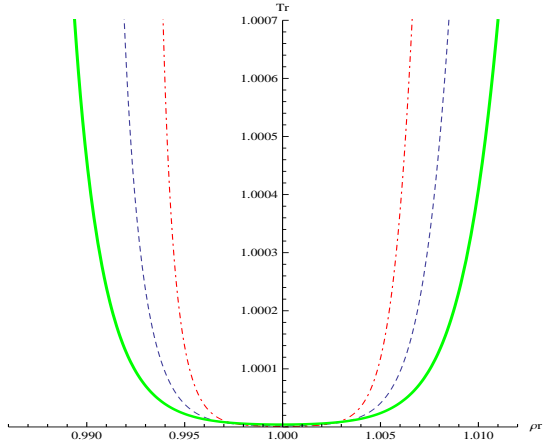}
\caption{Numerical solution for geodesics of the VdW equation of state close to criticality, in the $\left(\rho_r,T_r\right)$ plane. 
The dashed blue, dot-dashed red and solid green curves correspond to the boundary conditions 
$(T_r, \rho_r, {\dot \rho_r}) = \left(1.001, 1.009, -1.2\right)$, $\left(1.001, 1.007, -0.92\right)$, and $\left(1.0007, 1.011, -2.2\right)$ respectively.
The geodesics turn back from the critical point,
$\left(\rho_r,T_r\right) = \left(1,1\right)$.}
\label{vdwcriticalexpanded}
\end{figure}

Since we are interested in geodesics close to criticality (for a recent related discussion, see \cite{quevedo}), 
we now expand the metric upto first order about the critical point, $\left(T_r,\rho_r\right) = \left(1,1\right)$ 
(remember we have set $\left(T_c,\rho_c\right) = \left(1,1\right)$). The metric components are then given by the simple expressions
\begin{equation}
g_{TT}^c = 3\rho_r\left(\frac{3}{2} - T_r\right),~~~~g_{\rho\rho}^c = \frac{9}{4}\left(T_r - 1\right)
\label{vcrit}
\end{equation}
where the superscript $c$ in eq.(\ref{vcrit}) signifies that these expressions are valid close to criticality. The geodesic equations
of eq.(\ref{geodesic}) turn out to be
\begin{eqnarray}
&~&{\ddot T_r} + \frac{{\dot T_r}^2}{2T_r - 3} + \frac{{\dot T_r}{\dot \rho_r}}{\rho_r} + \frac{3{\dot \rho_r}^2}{4\rho_r\left(2T_r - 3\right)} = 0\nonumber\\
&~&{\ddot \rho_r} + \frac{{\dot \rho_r}{\dot T_r}}{T_r - 1} + \frac{{\dot T_r}^2\left(2T_r - 3\right)}{3\left(T_r - 1\right)} = 0
\label{vdwcritical}
\end{eqnarray}
We now numerically solve eq.(\ref{vdwcritical}) with three boundary conditions :
$(T_r, \rho_r, {\dot \rho_r})$ = $\left(1.001, 1.009, -1.2\right)$, $\left(1.001, 1.007, -0.92\right)$, and $\left(1.0007, 1.011, -2.2\right)$. 
\footnote{The value of ${\dot T_r}$ is fixed from the normalization condition as alluded to before.} For all the three
cases, we solve eq.(\ref{vdwcritical}) for values of the affine parameter between zero and $0.0025$. The solution for $T_r$ and $\rho_r$ are then
parametrically plotted by varying the affine parameter. The result is shown in fig.(\ref{vdwcriticalexpanded}) in
the $\left(\rho_r,T_r\right)$ plane, where the dashed blue,  dot-dashed red and solid green curves correspond to the three boundary conditions described above, 
respectively. We see that the geodesic curves ``turn back'' from the critical point. As we will see, this is a generic feature for all second order phase transitions
studied in this paper. 
\begin{figure}[t!]
\centering
\includegraphics[width=2.8in,height=2.3in]{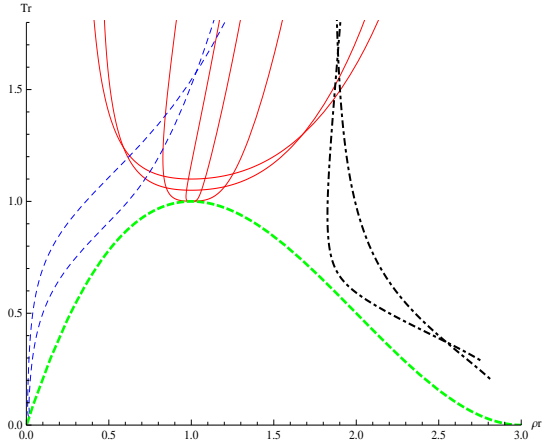}
\caption{Various numerical solution for geodesics of the VdW equation of state in the $\left(\rho_r,T_r\right)$ plane. The dashed blue, dot-dashed black and 
solid red lines are geodesics that begin from the gas, liquid and supercritical phases, respectively. The spinodal curve is shown in dotted green.}
\label{VDWfull}
\end{figure}


For the sake of completeness, we mention here that the analysis of geodesics using the full VdW metric of eq.(\ref{vdwfull}) is similar, although the 
geodesic equations are more complicated and we omit them for brevity. After extensive numerical analysis, our conclusion here is that 
a geodesic starting in the liquid ($\rho_r >1,T_r<1$) or gas ($\rho_r <1,T_r<1$) phase does not reach the other phase. They either terminate at the spinodal line or continue to the 
supercritical region. Also, close to the critical point, geodesics show the turn-around behavior as depicted in fig.(\ref{vdwcriticalexpanded}). We also find
that geodesics do not show any special behavior at the binodal lines, i.e at the location of the first order phase transitions, which is expected, because the
metric and the scalar curvature are both regular here. These results are summarized in fig.(\ref{VDWfull}), where we have shown several numerical solutions
to the geodesic equations for the VdW equation of state. The dotted green curve is the spinodal curve. The dashed blue curves on the left and the dot-dashed
black curves on the right are geodesics that start from the gas  and liquid phases respectively, and continue into the supercritical region. 
The solid red curves are geodesics in the super-critical region ($T_r > 1$), and show turning behaviour
similar to that depicted in fig.(\ref{vdwcriticalexpanded}).

We now move on to study geodesics in the classical mean-field Curie-Weiss ferromagnetic model in the thermodynamic limit. 
Information geometry of this model has been studied
extensively in \cite{drs}, and we simply state the result that in the $\left(T,m\right)$ representation, the line element of eq.(\ref{line}) is given by 
\begin{equation}
dl^2 = \frac{C_L}{T^2}dT^2 + \frac{1}{T}\frac{\left(T_c\left(1-m^2\right) -T\right)}{m^2-1} dm^2
\label{linecw}
\end{equation}
Here, $T$ is the temperature, $T_c$ its critical value, $m$ is the magnetization per unit spin, and 
$C_L(T)$ is a ``lattice specific heat'' introduced in \cite{jm},  that corresponds to the mechanical energy of the lattice. As was shown in \cite{jm}, 
information geometry in the CW model cannot be defined without introducing this term ad hoc in the theory. In \cite{drs}, it was shown that 
the line element in eq.(\ref{linecw}) correctly reproduces all the known features of the CW model, including the first order phase transitions. 
We will study the model close to criticality, and approximate the metric close to $m=0$ as 
\begin{equation}
g_{TT}^c = \frac{C_L(T)}{T^2},~~~~g_{mm}^c = 1 - \frac{T_c}{T}
\label{cwcric}
\end{equation}
where again the superscript $c$ denotes that we are close to criticality. 
To analyze the geodesic equations here, we note that a crucial simplification is possible, since none of the metric components in eq.(\ref{cwcric}) 
depend on the magnetization. This implies that the Lagrangian  of eq.(\ref{lagrangian}) is independent of $m$, and hence the Euler-Lagrange equation
that follows from it implies that ${\dot m} = K/g_{mm}^c$ where $K$ is a constant. Then from the normalization condition $g_{TT}^c{\dot T}^2 + g_{mm}^c{\dot m}^2 = 1$,
it follows that 
\begin{equation}
{\dot T}^2 = \frac{1}{g_{TT}^c}\left(1 - \frac{K^2}{g_{mm}^c}\right) =  \frac{T^2\left[T\left(1-K^2\right)-T_c\right]}{C_L(T)\left(T - T_c\right)}
\label{conditioncw}
\end{equation}
It is enough for us to consider the region $T>T_c$, for which eq.(\ref{conditioncw}) implies that
positivity of the right hand side imposes the restriction $T > T_c/(1-K^2)$, with $K^2 <1$. This means that
a geodesic in the region $T > T_c$ always remains in that region and cannot cross-over into the region $T< T_c$. A pathology arises for the case 
$K=0$, for which eq.(\ref{conditioncw}) implies that such a restriction is not implied, since ${\dot T}^2$ is always a positive number for $K=0$, or 
$m = {\rm constant}$. We have checked this by explicitly solving the geodesic equations, which in this case are given by
\begin{equation}
{\ddot T} + \frac{{\dot T}^2\left(T{\dot C_L} - 2C_L\right)}{2TC_L} - \frac{T_c{\dot m}^2}{2TC_L} = 0,~~~~{\ddot m} + \frac{T_c{\dot m}{\dot T}}{T\left(T - T_c\right)}=0
\label{geodesiccw}
\end{equation}
Numerical analyses (after choosing an appropriate regular functional form for $C_L(T)$, such as a power series) 
reveals that geodesics with $m = {\rm constant}$ lines (these are indeed geodesics as they satisfy the second equation
in eq.(\ref{geodesiccw})) cross over inside the spinodal region. This is probably a mathematical artifact and we
do not have a physical explanation for this. Apart from these constant $m$ lines, the behavior of geodesics close to the critical point 
is, as expected, qualitatively similar to that of the VdW fluid, and graphically, they resemble the ones shown in fig.(\ref{vdwcriticalexpanded}). 
We also find that the behavior of geodesics with the full CW metric (away from criticality) is qualitatively similar to those of the VdW model. Specifically,
geodesics in the phase $m > 0$ do not reach the phase $m < 0$, and vice versa.

\subsection{The Infinite Ising Ferromagnet}

We now study geodesics in the infinite-range ferromagnetic Ising model with a transverse magnetic field. 
This model was originally studied in \cite{diptiman}, where it was shown that
in the thermodynamic limit, it can be described by the classical dynamics of a single large spin. The information geometric
aspects of this model has not been studied so far, and we begin with a discussion on this. The Hamiltonian for this model is given by \cite{diptiman},\cite{amit}
\begin{equation}
H_{\rm IIF} = -\frac{J}{N}\sum_{i<j} S_i^zS_j^z - h\sum_iS_i^x = -\frac{J}{2N}\left(S_{\rm tot}^z\right)^2 - hS_{\rm tot}^x
\end{equation}
where the second equality follows from defining the total spin, $S_{\rm tot}^z = \sum_iS_i^z$, $S_{\rm tot}^x = \sum_iS_i^x$ (and neglecting a
constant term). We will set $J = 1$ in what follows. In a mean-field approach, where
the average magnetization $m = \sum_i<S_i^z>/N$, the Hamiltonian for a single spin reduces to $H_{\rm IIF}^1 = -mS_{\rm tot}^z - hS_{\rm tot}^x$. This is
an effective two-state model whose partition function can be shown to be given by 
\begin{equation}
Z = 2{\rm Cosh}\left(\frac{\sqrt{h^2 + m^2}}{2T}\right)
\end{equation}
To understand the geometric aspects of this model, we write the Gibbs free energy for the single spin, $G = -T{\rm ln} Z$ and effect a Legendre transform to obtain
the Helmholtz free energy $F = G + m^2/2$, where $m$ should be thought of as the applied magnetic field, i.e an intensive thermodynamic variable.  
The factor of $1/2$ in the Legendre transform might look strange, but note that this enforces the magnetization $\partial F/\partial m = 0$ 
(via the relation $m = -\partial G/\partial m$), i.e defines the boundary between the ferromagnetic and paramagnetic regions. In $(T,m)$ coordinates, 
using the expression for the Helmholtz free energy, the metric components are given from eq.(\ref{line}) by  
\begin{eqnarray}
g_{TT} &=& \frac{1}{4T^4}\left(h^2 + m^2\right){\rm Sech}^2\alpha \nonumber\\
g_{mm} &=& \frac{1}{T} - \frac{1}{4T^2}{\rm Sech}^2\alpha
\frac{\left(m^2\sqrt{h^2 + m^2} + h^2T{\rm Sinh}(2\alpha)\right)}{\left(h^2 + m^2\right)^{3/2}}
\label{iffmetric}
\end{eqnarray}
where $\alpha = \sqrt{h^2 + m^2}/2T$. 
\begin{figure}[t!]
\centering
\includegraphics[width=2.8in,height=2.3in]{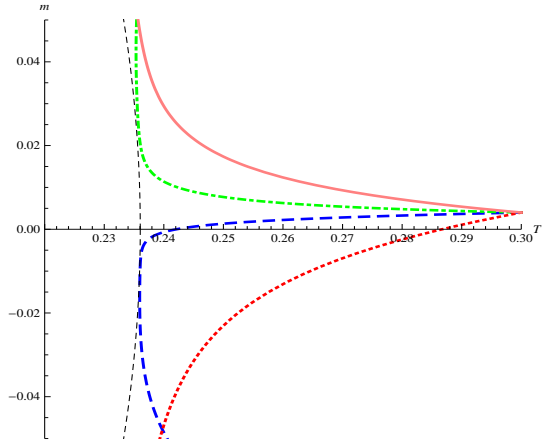}
\caption{Numerical solution for geodesics of the infinite range Ising ferromagnet at $h=0.2$, near the critical point $(T = 0.236, m=0)$. 
All geodesics are chosen to pass through the point $\left(T,m\right) = \left(0.3,0.004\right)$. The dashed blue, dot-dashed green, dotted red and solid 
pink curves correspond to the boundary condition ${\dot m} =$ $-0.03$, $0.03$, $-0.25$, and $0.12$ respectively. The dashed black line is the
spinodal curve, on which the scalar curvature diverges.}
\label{iif}
\end{figure}
The scalar curvature of eq.(\ref{scalarcurvature}) for the metric of eq.(\ref{iffmetric}), in the limit $m \to 0$ (which is our region of interest), is given by
$R = {\mathcal A}/{\mathcal B}$, where 
\begin{eqnarray}
{\mathcal A} &=& h \left[-2 T \left(4 h^2+4 T+1\right) {\rm Sinh} \left(\frac{h}{T}\right)+ 4 \left(h^2+2 T\right) {\rm Tanh}\left(\frac{h}{2 T}\right)
+3h {\rm Sech}^2\left(\frac{h}{2 T}\right)\right]
\nonumber\\
&-&2 T^2+2 h^2 (4T(T-2) -1)+2 T \left(4 h^2 (T+1)+T\right) {\rm Cosh} \left(\frac{h}{T}\right)\nonumber\\
{\mathcal B} &=& 2 h^2 \left({\rm Tanh} \left(\frac{h}{2 T}\right)-2 h\right)^2
\end{eqnarray}
The scalar curvature diverges at ${\rm Tanh}\frac{h}{2T} = 2h$, defining the phase boundary, a
result that matches with that obtained in \cite{diptiman}. To understand the behavior of geodesics in this model, we set $h=0.2$, which implies the critical
temperature $T = 0.236$. Numerical solutions of this geodesic equations close to the critical point 
are plotted in fig.(\ref{iif}). Here, we have taken all the geodesics to start
from $\left(T,m\right) = \left(0.3,0.004\right)$. The dashed blue, dot-dashed green, dotted red and solid pink curves correspond to 
${\dot m} =$ $-0.03$, $0.03$, $-0.25$, and $0.12$ respectively. Also shown in dashed black is the spinodal curve, i.e the locus of divergence of
the scalar curvature arising out of the metric of eq.(\ref{iffmetric}). We find that the geodesics show the same turning behavior as in the other
mean field models discussed in the previous subsection. 
We also note that in the limit of $T \to 0$, $g_{mm}$ diverges and $g_{TT} \to 0$. Numerical solutions seem to become somewhat 
unreliable in this limit, and we will not discuss them.

Having elucidated the nature of geodesics in classical systems exhibiting phase transitions at non-zero temperatures, we finally move to
quantum phase transitions at zero temperatures. 

\section{Geodesics in QPTs : The Transverse XY Spin Chain}

Information geometry of QPTs has been well studied of late, starting from the work of \cite{zan1}. There are, however, very few systems to which this
can be meaningfully applied, since the definition of the geometry (from eq.(\ref{metric})) 
requires complete knowledge of the many body ground state, which may be difficult to obtain excepting for a few exactly solvable system, like
the transverse field XY spin chain. Even when such ground states are obtainable, as in the Dicke model of quantum optics, explicit calculations might
be prohibitively difficult due to algebraic complications. We will base our calculations on the transverse XY model, for which the 
information metric was obtained in \cite{zan1}. 

To recall, for the transverse XY spin chain,  the Hamiltonian with $(2N + 1)$ spins is
\begin{equation}
H_{\rm XY} = -\left[\sum_{j = -N}^{N}\frac{1 + \gamma}{4}\sigma_j^x\sigma_{j+1}^x + \frac{1-\gamma}{4}\sigma_j^y\sigma_{j+1}^y - \frac{h}{2}\sigma_j^z\right]
\end{equation}
where the $\sigma^i$, $i=x,y,z$ are Pauli matrices, $\gamma$ is an anisotropy parameter, $h$ is the magnetic field, and the Planck's constant
has been set to unity.
The information metric for this model has been calculated in \cite{zan1} and in the thermodynamic limit, 
the line element, in the region $|h| <1$, $\gamma>0$ (the ferromagnetic phase) is given by
\begin{equation}
ds^2 = \frac{dh^2}{16\gamma\left(1-h^2\right)} + \frac{d\gamma^2}{16\gamma\left(1 + \gamma\right)^2}
\label{linexy}
\end{equation}
QPTs occur on the lines $\gamma=0, ~|h| \leq 1$ (the anisotropic transition line), and $|h| = 1$, (the Ising transition
lines), where the spectrum of the theory becomes gapless. Information geometry is however very different for these two transitions. Whereas the
scalar curvature (calculated from eqs.(\ref{scalarcurvature}) and (\ref{linexy})) diverges on the line $\gamma = 0$, it is regular on the lines $|h| = \pm 1$. 
For this model, the geodesic equations are 
\begin{equation}
{\ddot h} + \frac{h{\dot h}^2}{1-h^2} - \frac{{\dot h}{\dot \gamma}}{\gamma} = 0,~~{\ddot \gamma} - 
\frac{{\dot \gamma}^2\left(1+3\gamma\right)}{2\gamma\left(1 + \gamma\right)} + \frac{{\dot h}^2\left(1 + \gamma\right)^2}{2\gamma\left(1 - h^2\right)}=0
\label{geoxy}
\end{equation}
where, as before, the dot represents a derivative with respect to the affine parameter $\tau$. Also, the normalization 
condition implies that 
\begin{equation}
\frac{{\dot h}^2}{16\gamma\left(1 - h^2\right)} + \frac{{\dot \gamma}^2}{16\gamma\left(1 + \gamma\right)^2} = 1
\label{spacexy}
\end{equation}
\begin{figure}[t!]
\centering
\includegraphics[width=2.8in,height=2.3in]{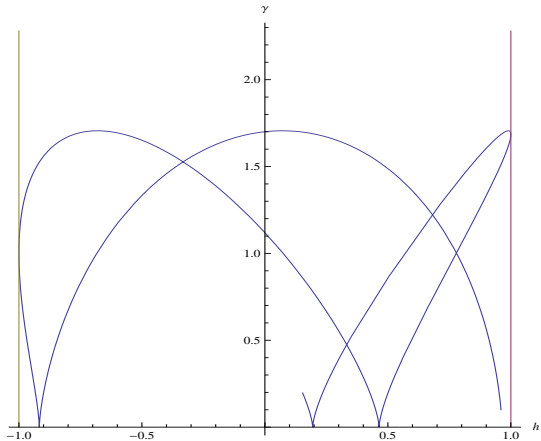}
\caption{Numerical solution for a geodesic curve with $(h,\gamma) = (0.96, 0.1)$ 
and $({\dot h},{\dot \gamma}) = (-0.0857,1.35)$ on the $h-\gamma$ plane. The geodesic is confined to a single phase region.}
\label{xy1}
\end{figure}
Before attempting to solve the coupled non-linear equations of eq.(\ref{geoxy}), let us look at a special case. The first of eq.(\ref{geoxy}) is satisfied
by $h = {\rm constant}$ and hence constant $h$ lines are geodesics. To find $\gamma$ as a function of the affine parameter in this case, we substitute
${\dot h} = 0$ in the second of eq.(\ref{geoxy}) and in eq.(\ref{spacexy}). Then it is seen that 
$\gamma = {\rm Tan}^2\left(2\left(\tau - \tau_0\right)\right)$, where $\tau_0$ is a reference value for the affine parameter. 
Thus, for the constant $h$ geodesics, $\gamma$ is always positive, i.e these geodesics 
do not cross the phase boundary at $\gamma = 0$. Rather, they turn back on touching that line. This should be 
contrasted with the $m = {\rm constant}$ geodesics of the CW model, which, as we have said is not fully understood. 

To solve the equations in eq.(\ref{geoxy}) in general, we adopt a numerical procedure analogous to what we have done before. As an illustration,
we solve for these equations with the initial conditions $(h,\gamma,{\dot h}) = (0.96, 0.1,-0.0857)$. The solution, plotted on the $h-\gamma$ plane 
parametrically, with the affine parameter $\tau$, is shown in fig.(\ref{xy1}). Clearly, the geodesic is confined to a single phase, and does not cross the 
phase boundaries, as in CPTs. It is not difficult to check this analytically by expanding the metric near the lines $\gamma = 0$ and $h = \pm 1$.

\section{Conclusions and Discussions}

In this paper, we have studied four model systems that exhibit phase transitions, in the thermodynamic limit. 
The Van der Waals model, the Curie-Weiss mean field model of
ferromagnetism and the infinite Ising ferromagnet exhibit CPTs at finite temperature. The transverse XY spin chain shows a QPT at zero temperature. 
For all these models, we performed the most general analysis of geodesics in the parameter manifold. Such an analysis has not appeared in the
literature before. In the process, we have established the information geometry of the infinite Ising ferromagnet. 
We have solved the geodesic equations for all these models in full generality, by choosing a starting point (i.e coordinates) in the manifold, and imposing 
initial conditions on its derivatives with respect to the affine parameter. In this way, we are able to trace out the geodesics, and study their behavior near second order
critical points. This complements and extends the results of \cite{diosi},\cite{zan2} in a non-trivial way. 

Our main conclusion here is that purely from a geometric perspective, geodesics near critical points show universal behavior, although the physical nature of 
the phase transitions are widely different. We have also established that geodesics are confined to a single phase. We believe that these
results are model independent, and should be true for any model of CPTs or QPTs. 

It might be interesting to study geodesics in the context of scaled equations of state for classical fluid systems, and also for some other models that exhibit QPTs. 
In particular, in the context of CPTs, it is an interesting question to ask if geodesics show any special behavior at or near the Widom line, which is a continuation
of the co-existence curve, along which the correlation length maximizes. We leave such a study for the future. 

\begin{center}
{\bf Acknowledgements}
\end{center}

It is a pleasure to thank Diptiman Sen for very useful correspondence. The work of SM is supported by grant no. 09/092(0792)-2011-EMR-1, from CSIR, India.\\

\end{document}